\documentclass[11pt,a4paper]{article}
\usepackage{jheppub}
\usepackage{amsmath}
\usepackage{amssymb}

\title{A Riccati type PDE for \\light-front higher helicity vertices}

\author[a,1]{Anders K. H. Bengtsson\note{Work supported by the Research and Education Board at the University of Bor{\aa}s.}}

\affiliation[a]{School of Engineering, University of Bor{\aa}s, All\' egatan 1, SE-50190 Bor\aa s, Sweden.}

\emailAdd{anders.bengtsson@hb.se} 

\abstract{This paper is based on a curious observation about an equation related to the tracelessness constraints of higher spin gauge fields. A similar equation also occurs in the theory of continuous spin representations of the Poincar{\'e} group. Expressed in an oscillator basis for the higher spin fields, the equation becomes a non-linear partial differential operator of the Riccati type acting on the vertex functions. The consequences of the equation for the cubic vertex is investigated in the light-front formulation of higher spin theory. The vertex is fixed by the PDE up to a set of terms that can be considered as boundary data for the PDE. These terms can serve as off-shell quantum corrections. 

In order to set the present work in perspective, some comments and comparisons to recent research on higher spin interactions are made. A few particular cubic vertices are calculated explicitly and compared to similar results in the literature, in particular the interesting cases $2-3-3$ and $3-2-2$ involving spin $2$ fields.}


\begin{document}

\maketitle

\pagebreak

\pagebreak

\section{Introduction}\label{sec:Introduction}
The structure of light-front higher spin cubic interactions has been investigated by Metsaev in a series of papers, in particular in \cite{Metsaev2005ar,Metsaev2007fb}. These results are for arbitrary dimensions and mixed symmetry fields. The general cubic vertex in four dimensions was considered by the present author in \cite{AKHB2012a}. That paper reviews and extends the original results of \cite{BBB1983a} and \cite{BBL1987}. It was found that the restrictions on the allowed vertices from the dynamical part of the Poincar{\'e} algebra are very weak. There is a countable infinite sequence of vertex operators in powers of oscillators and transverse momenta. This fact does not seem to be widely known but it is alluded to in \cite{Metsaev1993a}. One way of phrasing this result is to say that the light-front gauge-fixed cubic higher spin dynamics are (so far) restricted only by Poincar{\'e} kinematics: light-front cubic dynamics is Poincar{\'e} kinematically consistent. The non-abelian gauge algebra is not probed (as is well-known). In retrospect this was perhaps to be expected and it corroborates results of Metsaev \cite{Metsaev1993a} in four dimensions. However, when the light-front higher spin cubic vertices were first found in 1983, the result was indeed non-trivial given the state of knowledge at that time. Only no-go results were known and the existence of higher spin cubic vertices -- naturally generalizing spin 1 and 2 cubic interactions -- were quite unexpected. This was also the case for the Berends, Burgers and van Dam (BBvD) cubic covariant spin 3 self-interaction \cite{BerendsBurgersvanDam1984} published in 1984. 

Since then a huge amount of research has been done on higher spin interactions in flat space, in particular as regards cubic interactions. There are also partial results on quartic interactions \cite{Taronna2011a,DempsterTsulaia2012a}. Still, in the present author's opinion, the situation is not very clear. Some classic no-go theorems \cite{WeinbergS1964a,WeinbergWitten1980,ColemanMandula1967} put up severe obstacles to the consistency of interacting higher spin fields in Minkowski space. Furthermore, the very existence of the Vasiliev theory for interacting higher spin in AdS (see \cite{Vasiliev2004Reviewb,BekaertCnockaertIazeollaVasiliev2005} for reviews and references) is often put forward as evidence that such a theory does not exist in flat space. However, research into Minkowski space higher spin theory has been conducted in spite of such negative general theorems and quite a few yes-go examples have accumulated as reviewed in \cite{BekaertBoulangerSundellYesGo} (this paper also contains a list of more recent no-go results as well the yes-go examples).

In the first paper \cite{AKHB1988} on the covariant BRST-approach to interacting higher spin gauge fields in Minkowski spacetime, it was conjectured that the cubic vertex, apart from being gauge (BRST) invariant, also satisfies a special partial differential equation. This was shown for the pieces of the vertex then known and to the order that could be checked. Although the evidence is weak, the idea itself is quite intriguing, in particular as it can be connected to physical arguments. These arguments were set out in \cite{AKHB2013a} and will be repeated below. Since the light-front cubic vertex is known in its entirety, it is interesting to check if it satisfies a similar equation. This is the aim of the present paper.

It is found that the cubic vertex indeed satisfies a partial differential equation of the Riccati type. The coefficients of the vertex operator terms are related is such a way as to fix the vertex up to terms that can be considered as boundary data for the PDE. This freedom leave room for quantum corrections in the form of possible off-shell counter\-terms, although as will be discussed, the distinction between classical interaction terms and counterterms is not at all clear in a higher spin theory.

In section \ref{sec:LightFrontCubicVertex} the results from \cite{AKHB2012a} that we need are summarized. We also review (in section \ref{subsec:GeneralAnalysisCubInt}) the list of all cubic interaction terms from \cite{BBL1987} and explicitly calculate some interesting particular cases (in section \ref{subsec:Spin2Examples}) in order to compare to results from the recent literature. In section \ref{subsec:RelCurrentResearch} we briefly comment on the relation to results derived in covariant settings.
 Section \ref{sec:ThePDE} sets up and motivates the PDE. In section \ref{sec:Consequences} the consequences of the PDE are calculated. Section \ref{sec:Conclusion} contains conclusions and outlook. Conventions used are the same as in \cite{AKHB2012a} and the basic ones are listed in section \ref{sec:Conventions}.

\section{The general light-front cubic vertex in four dimensions}\label{sec:LightFrontCubicVertex}
The framework we are working in is spelled out in full in \cite{AKHB2012a}. Here we will only repeat the bare minimum needed for the present calculation. To begin with, fields of all helicities are collected in a Fock-space field
\begin{equation}\label{eq:FockSpaceField}
\vert\Phi(p)\rangle=\sum_{\lambda=0}^\infty\frac{1}{\sqrt{\lambda!}}\left(\phi_\lambda(p)(\bar\alpha^\dagger)^\lambda+\bar{\phi}_\lambda(p)(\alpha^\dagger)^\lambda\right)\vert 0\rangle.
\end{equation}
We are using a complex notation where $\phi_\lambda(p)$ is a field of helicity $+\lambda$ and its complex conjugate $\bar{\phi}_\lambda(p)$ has helicity $-\lambda$. Complex transverse momenta are denoted by $p$ and $\bar p$. The Fock field satisfies the constraint $\bar\alpha\alpha\vert\Phi(p)\rangle=0$ which is a remnant of the covariant theory tracelessness constraint. The full spectrum of Fock component fields is detailed in \cite{AKHB2012a}.

The free theory Hamiltonian is
\begin{equation}\label{eq:FreeHamiltonian}
H_{\scriptsize{(0)}}=\frac{1}{2}\int\gamma d\gamma dpd\bar{p}\langle\Phi\vert h\vert\Phi\rangle.
\end{equation}
where $\gamma$ denotes the $p^+$ component of momentum, and $h$ is $p\bar p/\gamma$. The cubic interaction can be written as
\begin{equation}\label{eq:CubicInteraction}
H_{\scriptsize{(1)}}=\frac{1}{3}\int\prod_{r=1}^3\gamma_rd\gamma_r dp_rd\bar{p}_r\langle\Phi_r\vert V_{123}\rangle.
\end{equation}
where the product over $r$ is over field enumeration $1,2,3$. The cubic vertex operator $\vert V_{123}\rangle$ as calculated in \cite{BBL1987} can be written
\begin{align}\label{eq:CubicVertex}\nonumber
\vert V_{123}\rangle&=\frac{g}{\kappa}\exp(\Delta_{\rm{st}}+\Delta_{\rm{hs}})\vert0_{123}\rangle\Gamma^{-1}\delta({\textstyle\sum_r}\gamma_r)\delta({\textstyle\sum_r}p_r)\delta({\textstyle\sum_r}\bar{p}_r)\\
&=\exp(\Delta_{\rm{st}}+\Delta_{\rm{hs}})|\varnothing_{123}\rangle
\end{align}
with
\begin{align}\label{eq:CubicVertexDeltasST}
\Delta_{\rm{st}}&=\sum_{r,s}Y^{rs}\alpha_r^\dagger\bar{\alpha}_s^\dagger+\kappa\sum_{r}Y^{r}(\alpha_r^\dagger\bar{\mathbb{P}}+\bar{\alpha}_r^\dagger\mathbb{P}),\\\label{eq:CubicVertexDeltasHS}
\Delta_{\rm{hs}}&=\kappa\sum_{r,s,t}Y^{rst}(\alpha_r^\dagger\alpha_s^\dagger\bar{\alpha}_t^\dagger\bar{\mathbb{P}}+\bar{\alpha}_r^\dagger\bar{\alpha}_s^\dagger\alpha_t^\dagger\mathbb{P}).
\end{align}
We use $|\varnothing_{123}\rangle$ as a shorthand for the Fock space vacua and the momentum conservation delta-functions and normalizing factors. The transverse momentum combination $\mathbb{P}$ is defined by 
\begin{equation}\label{eq:DefBlackboardMomentum}
\mathbb{P}=-\frac{1}{3}\sum_{r=1}^3\widetilde\gamma_r p_r\quad\text{with}\quad \widetilde\gamma_r=\gamma_{r+1}-\gamma_{r+2}
\end{equation}
and correspondingly for $\mathbb{\bar{P}}$.

In the formula \eqref{eq:CubicVertex}, $g$ is a dimensionless coupling constant and $\kappa$ is of dimension $-1$. For the cubic interactions $g$ becomes the spin-1 coupling and $g\kappa$ becomes the spin-2 coupling. The higher spin coupling constants $g_\lambda$ come out as $g\kappa^{\lambda-1}$. $\Gamma$ is $\gamma_1\gamma_2\gamma_3$ and it compensates the measure factor in \eqref{eq:CubicInteraction}.

The $Y^{r}$, $Y^{rs}$ and $Y^{rst}$ in formulas \eqref{eq:CubicVertexDeltasST} and \eqref{eq:CubicVertexDeltasHS} are rational functions of $\gamma$ that are determined by the Poincar{\'e} algebra. In \eqref{eq:CubicVertexDeltasST} we have
\begin{equation}\label{eq:CubicST}
Y^{rs}=\delta_{rs},\quad\quad\quad Y^{r}=\frac{1}{\gamma_r},
\end{equation}
and in \eqref{eq:CubicVertexDeltasHS}
\begin{equation}\label{eq:CubicHS}
Y^{rst}=\frac{\gamma_t}{\gamma_r\gamma_s}.
\end{equation}
The form of the string-like functions \eqref{eq:CubicST} are such that they cannot produce any self-interactions among the fields. For that the operator in \eqref{eq:CubicVertexDeltasHS} with $Y$-functions of the form \eqref{eq:CubicHS} (as discovered in \cite{BBL1987}) is needed.

The general vertices as derived in \cite{AKHB2012a} can be systematically listed using a shorthand notation for products of oscillators
\begin{equation}\label{eq:AlphaAbbreviations}
\boldsymbol \alpha_{r_k}^\dagger=\alpha_{r_1}^\dagger\ldots\alpha_{r_k}^\dagger\quad\text{ and }\quad\bar{\boldsymbol \alpha}_{s_l}^\dagger=\bar\alpha_{s_1}^\dagger\ldots\bar\alpha_{s_l}^\dagger,
\end{equation}
as
\begin{equation}\label{eq:VertexFunctionsGeneral}
\Delta=\kappa^{(n+m)}Y^{r_1\ldots r_ks_1\ldots s_l}\big(\boldsymbol \alpha_{r_k}^\dagger\bar{\boldsymbol \alpha}_{s_l}^\dagger\mathbb{P}^m\mathbb{\bar{P}}^n+\bar{\boldsymbol \alpha}_{r_k}^\dagger\boldsymbol \alpha_{s_l}^\dagger\mathbb{\bar{P}}^m\mathbb{P}^n\big).
\end{equation}
The functions are enumerated with $k=1,2,3,\ldots$ distinguishing two cases: $l<k$ and $l=k$. For the first case $l<k$ we get
\begin{equation}\label{eq:YFunctionsCase1}
Y^{r_1\ldots r_ks_1\ldots s_l}=\left(\frac{\gamma_{s_1}\ldots\gamma_{s_l}}{\gamma_{r_1}\ldots\gamma_{r_k}}\right)^{\frac{n+m}{n-m}},
\end{equation}
with $n-m=k-l$. Taking $m=0$ yields terms with either $\mathbb{P}$ or $\mathbb{\bar{P}}$. With non-zero $m$ we get terms with products of $\mathbb{P}$ or $\mathbb{\bar{P}}$. Such terms correspond to on-shell field redefinition terms. Hence, the case $m=0$ is the most interesting.

For the second case $l=k$ we get
\begin{equation}\label{eq:YFunctionsCase2}
Y^{r_1\ldots r_ks_1\ldots s_k}=\delta_{r_1s_1}\delta_{r_2s_2}\cdot\ldots\cdot\delta_{r_ks_k}.
\end{equation}
Excluding field redefinition terms we thus have the vertex functions
\begin{align}\label{eq:YFunctionsBasicSet}
Y^{r_1\ldots r_ks_1\ldots s_l}&=\frac{\gamma_{s_1}\ldots\gamma_{s_l}}{\gamma_{r_1}\ldots\gamma_{r_k}} &\text{ for } l<k,\\
Y^{r_1\ldots r_ks_1\ldots s_k}&=\delta_{r_1s_1}\delta_{r_2s_2}\cdot\ldots\cdot\delta_{r_ks_k} &\text{ for } l=k,
\end{align}
listed by enumerating $k=0, 1,2,\ldots$ and $0\leq l \leq k$. These are the functions we will work with in this paper. To be explicit, the cubic vertex operator we are considering is
\begin{align}\label{eq:VertexOperators}\nonumber
\Delta&=\sum_{\stackrel{k=1}{l\leq k}}\kappa^{(k-l)}y_{kl}Y^{r_1\ldots r_ks_1\ldots s_l}\big(\boldsymbol \alpha_{r_k}^\dagger\bar{\boldsymbol \alpha}_{s_l}^\dagger\mathbb{\bar{P}}^{k-l}+\bar{\boldsymbol \alpha}_{r_k}^\dagger\boldsymbol \alpha_{s_l}^\dagger\mathbb{P}^{k-l}\big)\\\nonumber
&=\sum_{\stackrel{k=1}{l<k}}\kappa^{(k-l)}y_{kl}\frac{\gamma_{s_1}\ldots\gamma_{s_l}}{\gamma_{r_1}\ldots\gamma_{r_k}}\big(\boldsymbol \alpha_{r_k}^\dagger\bar{\boldsymbol \alpha}_{s_l}^\dagger\mathbb{\bar{P}}^{k-l}+\bar{\boldsymbol \alpha}_{r_k}^\dagger\boldsymbol \alpha_{s_l}^\dagger\mathbb{P}^{k-l}\big)+\\
&\;\;\;\;\sum_{k=1}y_{kk}\delta_{r_1s_1}\delta_{r_2s_2}\cdot\ldots\cdot\delta_{r_ks_k}\big(\boldsymbol \alpha_{r_k}^\dagger\bar{\boldsymbol \alpha}_{s_k}^\dagger+\bar{\boldsymbol \alpha}_{r_k}^\dagger\boldsymbol \alpha_{s_k}^\dagger\big).
\end{align}
Since the relative coefficients for the terms are not fixed by the Poincar{\'e} algebra we have inserted numerical coefficients $y_{kl}$. The terms with $y_{kk}$ coefficients cannot contribute to any interactions. 

\paragraph{A note on covariantization}
One further comment is in order. The operators in \eqref{eq:VertexOperators} must in principle come from covariant operators. Not considering ghost contributions, that means operators of the form $\alpha_r\cdot\alpha_s$ and $\alpha_t\cdot p_u$ which would translate to light-front operators $\alpha_r^\dagger\bar\alpha_s^\dagger+\bar\alpha_r^\dagger\alpha_s^\dagger$ and $\alpha_r^\dagger\mathbb{\bar{P}}+\bar\alpha_r^\dagger\mathbb{P}$. Even more generally one could consider light-front operators constructed not just with the transverse metric $\eta_{ij}$ (where $i$ and $j$ runs over $1,2$) but also the transverse anti-symmetric tensor $\epsilon_{ij}$. In the complex notation used here that corresponds to working with independent operators $\alpha_r^\dagger\bar\alpha_s^\dagger$, $\bar\alpha_r^\dagger\alpha_s^\dagger$, $\alpha_r^\dagger\mathbb{\bar{P}}$ and $\bar\alpha_r^\dagger\mathbb{P}$. Then let
\begin{equation}\label{eq:MetsaevRepresentation}
X^{rs}=\frac{\gamma_s}{\gamma_r}\alpha_r^\dagger\bar\alpha_s^\dagger,\quad \bar X^{rs}=\frac{\gamma_s}{\gamma_r}\bar\alpha_r^\dagger\alpha_s^\dagger,\quad X^{r}=\frac{\alpha_r^\dagger\mathbb{\bar{P}}}{\gamma_r},\quad \bar X^{r}=\frac{\bar\alpha_r^\dagger\mathbb{P}}{\gamma_r}.
\end{equation}
as in \cite{Metsaev1993a}. In terms of these objects, the general vertex operator can be written as
\begin{equation}
x_{kl}X^{r_1s_1}X^{r_2s_2}\cdot\ldots\cdot X^{r_ks_k}X_{s_{k+1}}X_{s_{k+2}}\cdot\ldots\cdot X_{s_{k+l}}
\end{equation}
where $x_{kl}$ are numerical coefficients in an expansion over $k$ and $l$. To get a real expression one must add the complex conjugate. This representation, given by Metsaev in \cite{Metsaev1993a}, is equivalent to the representation used here.

\subsection{Classical interaction terms and counterterms}\label{subsec:ClassicalIntCount}
It must be realized that the vertex operators in \eqref{eq:VertexOperators} generate not just the pure helicity $\lambda$\,-$\lambda$\,-$\lambda$ classical interaction terms but also a large set of -- indeed all possible -- interactions between fields of different helicities. It also generates what could be considered as higher-derivative counterterms. However, as we will discuss further in section \ref{subsec:OffshellCounterterms}, classifying vertices as classical and counter-term respectively, is not at all un-ambiguous in a higher-derivative all-helicity theory. Let us here consider pure helicity $\lambda$ interactions, in particular $\lambda=2$. 

Suppose we want cubic interaction terms for helicity $\lambda$. The terms in the action are of the form $\phi_1\phi_2\phi_3$, $\bar\phi_1\phi_2\phi_3$, $\bar\phi_1\bar\phi_2\phi_3$ or $\bar\phi_1\bar\phi_2\bar\phi_3$ where the last two can be gotten from the first two using complex conjugation and cyclic symmetry in field labels. To extract an interaction of the form $\bar\phi_1\phi_2\phi_3$, we chose a bra state excited by $\bar\phi_1\phi_2\phi_3\alpha^{\lambda}_1\bar\alpha^{\lambda}_2\bar\alpha^{\lambda}_3$ and insert it into \eqref{eq:CubicInteraction} using \eqref{eq:VertexOperators}. The matrix element to compute is
\begin{equation}\label{eq:BarPhiPhiPhi}
\bar\phi_1\phi_2\phi_3\langle0\vert\alpha_1^{\lambda}\bar\alpha_2^{\lambda}\bar\alpha_3^{\lambda}e^\Delta\vert\varnothing_{123}\rangle.
\end{equation}
The annihilators will saturate any combination of creators $({\bar\alpha_1}^\dagger)^{\lambda}(\alpha_2^\dagger)^{\lambda}(\alpha_3^\dagger)^{\lambda}$ that appear in the expansion of $\exp(\Delta)$. This will give the familiar basic higher spin interactions with minimal number of transverse derivatives equal to the helicity $\lambda$. But it will also pick out higher derivate terms. Some of these can be interpreted as counterterms. For instance, in the case of helicity $2$ we will get various terms with transverse momentum factors of $\mathbb{\bar{P}}^3\mathbb{P}$ and $\mathbb{\bar{P}}^4\mathbb{P}^2$ corresponding to one loop and two loops respectively. 

On the other hand, to extract an interaction of the form $\phi_1\phi_2\phi_3$, we chose a bra state excited by $\phi_1\phi_2\phi_3\bar\alpha^{\lambda}_1\bar\alpha^{\lambda}_2\bar\alpha^{\lambda}_3$ and insert it into \eqref{eq:CubicInteraction} using \eqref{eq:VertexOperators}. The matrix element to compute is
\begin{equation}\label{eq:PhiPhiPhi}
\phi_1\phi_2\phi_3\langle0\vert\bar\alpha_1^{\lambda}\bar\alpha_2^{\lambda}\bar\alpha_3^{\lambda}e^\Delta\vert\varnothing_{123}\rangle.
\end{equation}
For helicity $2$ this will produce an off-shell two loop counter term with transverse derivative structure $\mathbb{\bar{P}}^6$ \cite{AKHB2013b}. A few more comments on counterterms involving higher spin will be given in sections \ref{subsec:OffshellCounterterms} and \ref{subsec:GravityCounterterms}.

\subsection{General analysis of cubic interactions}\label{subsec:GeneralAnalysisCubInt}
The overall structure of the possible interaction terms can be analyzed in the following way. First again consider an interaction of the form $\bar\phi_1\phi_2\phi_3$, but now with fields of three -- possibly different -- helicities. Chose a bra state excited by $\bar\phi_{\lambda_1}\phi_{\lambda_2}\phi_{\lambda_3}\alpha^{\lambda_1}_1\bar\alpha^{\lambda_2}_2\bar\alpha^{\lambda_3}_3$. This saturates an operator combination $({\bar\alpha_1}^\dagger)^{\lambda_1}(\alpha_2^\dagger)^{\lambda_2}(\alpha_3^\dagger)^{\lambda_3}\mathbb{\bar{P}}^n\mathbb{P}^m$ from the exponential $\exp\Delta$. We then get the condition $n-m=\lambda_2+\lambda_3-\lambda_1$. Taking $m=0$ we get interactions with pure powers of $\mathbb{\bar{P}}^n$ where $n=\lambda_2+\lambda_3-\lambda_1\geq0$, that is with $\lambda_2+\lambda_3\geq\lambda_1$. This corresponds to the terms classified as (ii) in \cite{BBL1987} (see formula A1.4). On the other hand, taking $n=0$ we get interactions with pure powers of $\mathbb{P}^m$ where $m=\lambda_1-\lambda_2-\lambda_3\geq0$, that is with $\lambda_1\geq\lambda_2+\lambda_3$. This corresponds to the terms classified as (iii) in \cite{BBL1987} (see formula A1.5). The case (i) is included with $\lambda_1=\lambda_2+\lambda_3$.

Higher derivative, pure helicity $\lambda$ on-shell interaction terms (generalizing the $\lambda=2$ on-shell counterterms) are produced if we take $\lambda_1=\lambda_2=\lambda_3=\lambda$. Then $n-m=\lambda$. The maximum value of $m$ is $\lambda$ (there are just $\lambda$ oscillators $\alpha$ in the state), we get a sequence of interaction terms with transverse momentum structure $\mathbb{\bar{P}}^\lambda, \mathbb{\bar{P}}^{\lambda+1}\mathbb{P}, \mathbb{\bar{P}}^{\lambda+2}\mathbb{P}^2, \ldots, \mathbb{\bar{P}}^{2\lambda}\mathbb{P}^\lambda$. We recognize the helicity $2$ sequence $\mathbb{\bar{P}}^2, \mathbb{\bar{P}}^3\mathbb{P}, \mathbb{\bar{P}}^4\mathbb{P}^2$ corresponding to zero, one and two loops. For higher helicity $\lambda$, the match with pure $\lambda$ cubic divergencies is not so exact. For instance for $\lambda=3$, where the sequence runs $\mathbb{\bar{P}}^3, \mathbb{\bar{P}}^4\mathbb{P}, \mathbb{\bar{P}}^5\mathbb{P}^2, \mathbb{\bar{P}}^6\mathbb{P}^3$, only the $\mathbb{\bar{P}}^5\mathbb{P}^2$ term matches a one-loop pure spin $3$ cubic divergence of degree $7$.

Next consider interactions excited by $\phi_{\lambda_1}\phi_{\lambda_2}\phi_{\lambda_3}\bar\alpha^{\lambda_1}_1\bar\alpha^{\lambda_2}_2\bar\alpha^{\lambda_3}_3$. In this case we simply get interactions with transverse momentum structure $\mathbb{\bar{P}}^m$ where $m=\lambda_1+\lambda_2+\lambda_3$. This corresponds to the terms classified as (iv) in \cite{BBL1987} (see formula A1.6).

The interactions with field combinations $\phi_1\bar\phi_2\bar\phi_3$ and $\bar\phi_1\bar\phi_2\bar\phi_3$ should of course be added for hermiticity, but they are gotten by complex conjugation and cyclic symmetry in field labels. In this way we retrieve all possible cubic interactions listed in our 1987 paper \cite{BBL1987}. 
For instance the pure $\lambda-\lambda-\lambda$ self-interaction vertices come out with momentum structure
\begin{equation}\label{eq:SelfInteraction}
\left(\frac{\gamma_1}{\gamma_2\gamma_3}\mathbb{\bar{P}}\right)^\lambda=(p_1)^\lambda\left(\frac{\bar p_2}{p_2^+}-\frac{\bar p_3}{p_3^+}\right)^\lambda,
\end{equation}
and we see that higher spin cubic self-interaction vertices are powers of the Yang-Mills cubic vertex. Transformed to configuration space these vertices correspond exactly to the cubic self-interactions found in  \cite{BBB1983a}. Let us now turn to examples with spin 2 interactions.

\subsection{Examples with spin 2 interactions}\label{subsec:Spin2Examples}
Higher spin interactions with gravity -- or spin $2$ -- is an important topic. Considering $s_1-s_2-s_3$ interactions there are two sub-cases to consider: $2-s-s$ and $s-2-2$ interactions. The first case is the one that is normally considered as gravitational interaction of a spin $s$ field and it can be thought of as a spin $s$ field propagating in a gravitational background. The case $s-2-2$ is more like a spin $2$ field propagating in a spin $s$ background.

Since we are working in a helicity basis it is convenient to denote a negative helicity by an overbar, say $\bar 2$ for $-2$. To avoid confusion, in writing $s$ we allow $s$ be negative, zero or positive. When writing $\lambda$ we always have $\lambda\geq0$ and we write $\bar{\lambda}=-\lambda$ for a negative helicity. In some of the vertices below the spin 2 field needs to be colored.

\paragraph{Gravitational $2-s-s$ interactions}
Start with interactions corresponding to case (ii) above, i.e. $\bar{\phi}\phi\phi$ interactions with powers of $\mathbb{\bar{P}}$. We then have two sub-cases: (ii-a) $\bar{2}-\lambda-\lambda$ and (ii-b) $2-\lambda-\bar{\lambda}$. In case (ii-a) the power $n$ of $\mathbb{\bar{P}}$ become $n=\lambda+\lambda-2=2(\lambda-1)$. Then $\lambda=2$ corresponds to the pure gravitational cubic vertex while for instance $\lambda=3$ with $n=4$ yields the recently covariantly studied higher derivative vertex \cite{BoulangerLeclerc2006a,BoulangerSundellLeclerc2008,SagnottiTaronna2011a} (for more comments see section \ref{subsec:RelCurrentResearch}). The light-front form of the $\bar{2}-3-3$ vertex has the momentum structure
\begin{equation}\label{eq:Grav2ssCaseiia1}
\frac{\gamma_1^2}{\gamma_2^3\gamma_3^3}\mathbb{\bar{P}}^4=\gamma_1^2\gamma_2\gamma_3\left(\frac{\mathbb{\bar{P}}}{\gamma_2\gamma_3}\right)^4=\gamma_1^2\gamma_2\gamma_3\left(\frac{\bar p_2}{\gamma_2}-\frac{\bar p_3}{\gamma_3}\right)^4,
\end{equation}
where the last equality follows from momentum conservation. The vertex Fourier transforms into the configuration space form
\begin{equation}\label{eq:Grav2ssCaseiia2}
\int d^3x\sum_{n=0}^4(-1)^n{4\choose n}(\partial^+)^2\bar\phi_{(2)}\Big[\frac{\bar\partial}{\partial^+}\Big]^{4-n}\partial^+\phi_{(3)}\Big[\frac{\bar\partial}{\partial^+}\Big]^{n}\partial^+\phi_{(3)},
\end{equation}
where $\phi_{(\lambda)}$ denotes a helicity $+\lambda$ field and $\bar\phi_{(\lambda)}$ denotes a helicity $-\lambda$ field. The color index $a$ (not explicitly shown) on the spin $3$ field is contracted as $\phi^a\phi^a$ consistent with the even power $4$ of transverse derivatives in the binomial expansion.

Turning to case (ii-b) the power $n$ of $\mathbb{\bar{P}}$ become $n=2+\lambda-\lambda=2$ independent on $\lambda$. These interactions are unexpected as they correspond to what normally would be called minimal coupling of higher spin to gravity. It is well known that such interactions are inconsistent by themselves as noted already by Fierz and Pauli in \cite{FierzPauli} and then investigated by many authors \cite{Buchdahl1962,JohnsonSudarshan,VeloZwanziger,AragoneDeser1971,AragoneDeser1980a,AragoneDeser1979,AragoneDeser1980b,BerendsHoltenWitNieuwenhuizen,BarthChristensen,AragoneLaRoche,Curtright1980}. A no-go theorem (Weinberg-Witten) was proved in \cite{WeinbergWitten1980} and further strengthened in \cite{Porrati2008}. The occurrence of this vertex here is a reminder that light-front cubic interactions do not probe the gauge algebra. As already pointed out based on the analysis in \cite{AKHB2012a}, closing the Poincar{\'e} algebra on the cubic level yields all possible cubic interactions compatible with kinematics. Metsaev in \cite{Metsaev2005ar} do not find these interactions in $D>4$ but according to \cite{Metsaev1993a} they should occur in four dimensions. The somewhat special properties of $D=4$ in higher spin theory need to be further studied. The explicit form of the $\bar 3-2-3$ vertex is
\begin{equation}\label{eq:Grav2ssCaseiib1}
\frac{\gamma_1^3}{\gamma_2^2\gamma_3^3}\mathbb{\bar{P}}^2=\frac{\gamma_1^3}{\gamma_3}\left(\frac{\mathbb{\bar{P}}}{\gamma_2\gamma_3}\right)^2=\frac{\gamma_1^3}{\gamma_3}\left(\frac{\bar p_2}{\gamma_2}-\frac{\bar p_3}{\gamma_3}\right)^2.
\end{equation}
The vertex Fourier transforms into the configuration space form
\begin{equation}\label{eq:Grav2ssCaseiib2}
\int d^3x\sum_{n=0}^2(-1)^n{2\choose n}(\partial^+)^3\bar\phi_{(3)}\Big[\frac{\bar\partial}{\partial^+}\Big]^{2-n}\phi_{(2)}\Big[\frac{\bar\partial}{\partial^+}\Big]^{n}\frac{1}{\partial^+}\phi_{(3)}.
\end{equation}

Next we consider interactions corresponding to case (iii) above, i.e. $\bar{\phi}\phi\phi$ interactions with powers of $\mathbb{P}$ transverse momenta. Again there are two sub-cases: (iii-a) $\bar{2}-\lambda-\lambda$ and (iii-b) $2-\lambda-\bar{\lambda}$. In case (iii-a) the power $m$ of $\mathbb{P}$ become $m=2-\lambda-\lambda=2(1-\lambda)$ and we must have $\lambda\leq1$. Then $\lambda=0$ gives the gravitational interaction of a scalar field. Case (iii-b) is empty since the power $m$ of $\mathbb{P}$ become $m=\lambda-2-\lambda=-2$.

Finally we consider interactions corresponding to case (iv) above, i.e. $\phi\phi\phi$ interactions with powers of $\mathbb{\bar{P}}$ transverse momenta. We have $m=2+\lambda+\lambda=2(1+\lambda)$. For spin 3 this yields a vertex with eight factors of transverse momenta $\mathbb{\bar{P}}^8$. A covariant version of such a vertex was found in \cite{SagnottiTaronna2011a}. The explicit form of the light-front $2-3-3$ vertex is
\begin{equation}\label{eq:Grav2ssCaseiv1}
\frac{\mathbb{\bar{P}}^8}{\gamma_1^2\gamma_2^3\gamma_3^3}=\frac{\gamma_2^5\gamma_3^5}{\gamma_1^2}\left(\frac{\mathbb{\bar{P}}}{\gamma_2\gamma_3}\right)^8=\frac{\gamma_2^5\gamma_3^5}{\gamma_1^2}\left(\frac{\bar p_2}{\gamma_2}-\frac{\bar p_3}{\gamma_3}\right)^8.
\end{equation}
The vertex Fourier transforms into the configuration space form
\begin{equation}\label{eq:Grav2ssCaseiv2}
\int d^3x\sum_{n=0}^8(-1)^n{8\choose n}\frac{1}{(\partial^+)^2}\phi_{(3)}\Big[\frac{\bar\partial}{\partial^+}\Big]^{8-n}(\partial^+)^5\phi_{(2)}\Big[\frac{\bar\partial}{\partial^+}\Big]^{n}(\partial^+)^5\phi_{(3)}.
\end{equation}

The $2-3-3$ vertex of \cite{SagnottiTaronna2011a} with six factors of transverse momenta can only reproduced here with transverse momentum structure $\mathbb{\bar{P}}^4\mathbb{P}^2$ and thus is on-shell. All in all we find interactions of the type $2-3-3$ with two, four, six and eight factors of transverse derivatives. This is in accordance with the work of Sagnotti and Taronna in \cite{SagnottiTaronna2011a} based on analyzing the cubic interactions following from the first Regge trajectory of the open string.
 
\paragraph{Spin $2$ in a higher spin background: $s-2-2$ interactions}The existence of this kind of light-front cubic interaction follows from the supersymmetric vertices of \cite{BBB1983b}. The covariant version is also mentioned in \cite{BerendsBurgersvanDam1985} and further analyzed in Burgers' thesis \cite{Burgers1985thesis}. 

Again start with interactions corresponding to case (ii) above, i.e. $\bar{\phi}\phi\phi$ interactions with powers of $\mathbb{\bar{P}}$ transverse momenta. We then have two sub-cases: (ii-a) $\bar{\lambda}-2-2$ and (ii-b) $\bar{2}-2-\lambda$. In case (ii-a) the power $n$ of $\mathbb{\bar{P}}$ become $n=2+2-\lambda=4-\lambda$.

Turning to case (ii-b) the power $n$ of $\mathbb{\bar{P}}$ become $n=2+\lambda-2=\lambda$. These interactions have recently been confirmed for $s=3$ in a covariant formalism \cite{BoulangerLeclerc2006a}. Note that the number of derivatives in the vertex is equal to the spin $s$. This case is interesting since it hints at interactions where it is the higher spin field that governs the interaction with spin $2$ rather than the other way around. The explicit form of the light-front $\bar 2-2-3$ vertex is
\begin{equation}\label{eq:HSs22Caseiib1}
\frac{\gamma_1^2}{\gamma_2^2\gamma_3^3}\mathbb{\bar{P}}^3=\gamma_1^2\gamma_2\left(\frac{\mathbb{\bar{P}}}{\gamma_2\gamma_3}\right)^3=\gamma_1^2\gamma_2\left(\frac{\bar p_2}{\gamma_2}-\frac{\bar p_3}{\gamma_3}\right)^3.
\end{equation}
The vertex Fourier transforms into the configuration space form
\begin{equation}\label{eq:HSs22Caseiib2}
\int d^3x\sum_{n=0}^3(-1)^n{3\choose n}(\partial^+)^2\phi_{(2)}\Big[\frac{\bar\partial}{\partial^+}\Big]^{3-n}\partial^+\phi_{(2)}\Big[\frac{\bar\partial}{\partial^+}\Big]^{n}\phi_{(3)}.
\end{equation}
The spin 2 field needs to be colored in this interaction.

Next we consider interactions corresponding to case (iii), i.e. $\bar{\phi}\phi\phi$ interactions with powers of $\mathbb{P}$ transverse momenta. The two sub-cases are (iii-a) $\bar{\lambda}-2-2$ and (iii-b) $\bar{2}-\lambda-2$. In the case (iii-a) we get $m=\lambda-2-2=\lambda-4$. In the case (iii-b) we get $m=2-\lambda-2=-\lambda$ so this case is empty.

In the last case (iv) we have $\lambda-2-2$ interactions. The power of transverse momenta $\mathbb{\bar{P}}$ is $\lambda+4$.
\paragraph{Conclusion}
Although the gravitational spin $2$ field still may play a special role as the metric field, higher spin theory tends to demote spin $2$ to just one field in the infinite spectrum of higher spin fields. If higher spin field theory eventually turns out to be fully consistent it is likely that there will be high energy - high spin corrections to gravity as we know it. The presence of these kinds of interactions is therefore an indication of a more egalitarian role of spin 2 at very high energies. No particular spin is singled out as special in higher spin theory. If the theory is consistent, then we have to think of any particular spin as propagating in a self-consistent background of all spins. One point of worry in that case is that it may not make much sense to work in terms of field theory, in particular in terms of field theory on a fixed background. That is however a much deeper question that cannot be addressed here.

\subsection{Relation to current research on Minkowski space interactions}\label{subsec:RelCurrentResearch}
Over the last ten years a large stock of knowledge on cubic interactions has been gathered by many authors. The picture that emerges is a general consensus on the overall structure of the cubic interactions. The results have been derived using different formalistic apparatuses: (i) field theoretic Noether-coupling/BRST methods (see for instance \cite{BarnichHenneaux1993a,BarnichBrandtHenneaux2000,HenneauxTeitelboim}) generalizing the approach of the 1984 work of Berends, Burgers  and vanDam \cite{BoulangerLeclerc2006a}, (ii) limits of the Vasiliev theory (see for instance \cite{FradkinVasiliev1987Cubic,FradkinVasiliev1987Grav}) cubic interactions \cite{BoulangerSundellLeclerc2008}, (iii) limits of string amplitudes \cite{SagnottiTaronna2011a,Polyakov2010a}, (iv) string field theory (see for instance \cite{Witten1986a,CBThorn1989,GrossJevicki1987II}) inspired BRST-methods \cite{BuchbinderFotopoulosPetkouTsulaia2006a,FotopoulosTsulaia2010a,DempsterTsulaia2012a} in the spirit of the present author's 1988 paper \cite{AKHB1988}, (v) FDA-$L_\infty$-methods \cite{Taronna2011a}, (vi) generating function Noether methods \cite{ManvelyanMkrtchyanRuhl2010a,ManvelyanMkrtchyanRuhl2011a} generalizing BBvD  \cite{BerendsBurgersvanDam1984,BerendsBurgersvanDam1985,Burgers1985thesis}, and finally (vii) light-front methods by Metsaev \cite{Metsaev1993a,Metsaev2005ar,Metsaev2007fb} extending the original results of \cite{BBB1983a,BBB1983b,BBL1987}.

The many different -- but related -- formalisms and choices regarding gauge, on-shell \& off-shell, field redefinitions, make detailed term-by-term comparisons cumbersome (and has not been exhaustively undertaken yet as it seems). A general feature that emerges from these works is the naturalness of introducing an auxiliary classical or first quantized variable, say $\xi_\mu$, parameterizing the higher spin gauge field symmetric tensors. This is a feature that goes back to Fronsdal's work of the 1970's \cite{Fronsdal1979conf} (for a review see \cite{AKHB2008a}). Whether this is just a convenient trick or there is some underlying physics behind it, remains to be uncovered. It can be argued that there ought to be physics behind it in order that massless higher spin fields (or whatever degrees of freedom replace fields at extreme high energies) play a role in fundamental physics. This is certainly the opinion of the present author.

We will not attempt a comprehensive review here, but rather make a few comments and observations complementing the comparisons already made in section \ref{subsec:Spin2Examples}. We warn the reader that what follows below is a very qualitative view of a technically demanding quantitative literature. We follow the classification and numbering (i) - (vi).

\begin{description}
  \item[(i)] 
In \cite{BoulangerLeclerc2006a}  the authors investigate couplings between spin 2 spin 3 fields. They find a non-minimal $2-3-3$ vertex with four derivatives. This corresponds to the light-front vertex in formula \eqref{eq:Grav2ssCaseiia2} above. They also find the $3-2-2$ vertex with three derivatives corresponding to spin 2 in a spin 3 background \cite{BerendsBurgersvanDam1985} (corresponding to light-front vertex in formula \eqref{eq:HSs22Caseiib1}). These results are then further studied in \cite{BoulangerSundellLeclerc2008} where the $2-3-3$ vertex is shown to be unique and consistent to the next order.

  \item[(ii)] 
Furthermore in \cite{BoulangerSundellLeclerc2008} the vertices of \cite{BoulangerLeclerc2006a} are compared to limits of the Vasiliev cubic interactions \cite{FradkinVasiliev1987Cubic,FradkinVasiliev1987Grav}. The vertex $2-3-3$ is the leading term in the flat limit. However there are also sub-leading terms, one of which corresponds to a two-derivative minimal gravitational coupling of spin 3 fields, an interaction that is ruled out by flat space no-go theorems, but are present in the light-front formulation.
  \item[(iii)] 
The results of (i) and (ii) are corroborated by the string theory analysis of \cite{SagnottiTaronna2011a}. Studying the first Regge trajectory, a sequence of $2-3-3$ vertices with powers of momentum factors $0,2,4,6,8$ is found. Since this analysis is based on string theory there naturally appears a variable $\xi$ related to the string theory oscillators. In terms of such variables, the interactions that survive in the massless limit are essentially generated by differential operators of the form
\begin{equation}\label{eq:SagnottiTaronnaOperator}
\mathcal{G}=\sqrt{\alpha^\prime/2}[(\partial_{\xi_1}\cdot\partial_{\xi_2})(\partial_{\xi_3}\cdot p_{12})+(\partial_{\xi_2}\cdot\partial_{\xi_3})(\partial_{\xi_1}\cdot p_{23})+(\partial_{\xi_3}\cdot\partial_{\xi_1})(\partial_{\xi_2}\cdot p_{31})]
\end{equation}
where $1,2,3$ indexes external states and where $p_{12}=p_1-p_2$ et cetera, in terms of external momenta. This is interesting because this form corresponds to the light-front operators $\Delta_{\mathrm{hs}}$ of \eqref{eq:CubicVertexDeltasHS} of the present work, and in a covariant framework to the vertex operators computed in \cite{AKHB1988}. See also item (vi) below.

  \item[(iv)]
For recent work on string field theory inspired BRST methods, see \cite{BuchbinderFotopoulosPetkouTsulaia2006a,FotopoulosTsulaia2010a,DempsterTsulaia2012a} and references therein.

  \item[(v)]
In \cite{Taronna2011a} Taronna tries to synthesize much of the accumulated knowledge of cubic (and quartic) interactions into a comprehensive scheme based on a general construction of FDA's (see for instance \cite{DAuriaFre1982a}) and $L_\infty$ methods in the spirit of \cite{Stasheff1997a,LadaStasheff1993a,FulpLadaStasheff2002,Zwiebach1993a}. See also \cite{AKHB2005a,AKHB2007a} for the present author's attempts along such lines. 

That there is a structure of homological algebra behind field theory as expressed in a Batalin-Vilkovisky language was first realized by Stasheff in \cite{Stasheff1997a}. This was later elaborated into a proof that if the BBvD theory is consistent, then the interactions and gauge transformations must form a $L_\infty$-algebra \cite{LadaStasheff1993a,FulpLadaStasheff2002}. Since the BBvD theory -- at least ostensibly -- is a single spin theory, it cannot be (and indeed is not) consistent. But it is quite clear from Zwiebach's work \cite{Zwiebach1993a} on closed string field theory that an all-spin theory -- if it is consistent -- falls into the structure of homological algebra.

  \item[(vi)] 
In \cite{ManvelyanMkrtchyanRuhl2010a,ManvelyanMkrtchyanRuhl2011a} the authors derive covariant cubic interactions for any combination of spin. They work with symmetric tensor fields expanded over a variable of the type $\xi_\mu$, satisfying the free Fronsdal equations. They work out a generating function for the interactions of the form \eqref{eq:SagnottiTaronnaOperator} supplemented by an operator of the general form
\begin{equation}\label{eq:ManvelyanMkrtchyanRuhlExtraOperator}
\sim(\partial_{\xi_1}p_{23}+\partial_{\xi_2}p_{31}+\partial_{\xi_3}p_{12})
\end{equation}
again confirming the overall structure of at least part of the covariant cubic vertex.

Thus we have evidence from at least four more or less independent sources that cubic higher spin interactions can be parameterized by generating functions of the qualitative form of formulas \eqref{eq:SagnottiTaronnaOperator} and \eqref{eq:ManvelyanMkrtchyanRuhlExtraOperator}: the light-front approach \cite{BBL1987,Metsaev1993a,Metsaev2005ar}, BRST approach \cite{AKHB1988,FotopoulosTsulaia2010a}, limits of string approach \cite{SagnottiTaronna2011a} and Noether approach \cite{ManvelyanMkrtchyanRuhl2010a,ManvelyanMkrtchyanRuhl2011a}.

\end{description}  

\paragraph{The question of minimal coupling}
However one issue remains here, at least in four dimensions, where the light-front vertex operators produce minimal gravitational $2-s-s$ and $1-s-s$ Yang-Mills couplings. This goes against the no-go theorems. One way out is to realize that a purely cubic analysis is not likely to probe deep enough to see such inconsistencies as the no-go theorems indicate, and that these interactions will indeed turn out to be impossible when higher order interactions are considered. However, that way out is not open if the PDE of the present work is taken seriously. Then minimal $2-s-s$ and $1-s-s$ interactions cannot be avoided as they are part of the full cubic vertex. Phrased in terms of variables used in the operator \eqref{eq:SagnottiTaronnaOperator} above, such parts of the vertex contain higher powers of $\partial_{\xi_1}\partial_{\xi_2}+\partial_{\xi_2}\partial_{\xi_3}+\partial_{\xi_3}\partial_{\xi_1}$. 

This is a conundrum and the only way to save the theory would be if the no-go theorems are not applicable to the fully interacting theory. That this is not an entirely unthinkable scenario is indicated by the fact that minimal interactions are implied both in AdS Vasiliev theory and in string theory \cite{BoulangerSundellLeclerc2008,SagnottiTaronna2011a}. It seems fair to say that the situation as regards cubic and quartic interactions in four dimensional flat space is not yet fully clarified.

\section{The PDE}\label{sec:ThePDE}
The explicit form of the differential equation is as follows
\begin{equation}\label{eq:UDequation0}
\sum_{r=1}^3\left((\bar{\alpha}_r\alpha_r-\bar{\alpha}^\dagger_r\alpha_r-\alpha^\dagger_r\bar{\alpha}_r)\Delta+(\bar{\alpha}_r\Delta)(\alpha_r\Delta)\right)=-\sum_{r=1}^3\bar{\alpha}^\dagger_r\alpha^\dagger_r+\rho^2\sum_{r=1}^3\frac{\bar{\mathbb{P}}\mathbb{P}}{\gamma_r^2}+3\eta.
\end{equation}
where the sum is over the three transverse Fock-spaces connected by the cubic vertex, suggesting a generalization to higher order vertices. All the terms (in particular the oscillators) in the equation are acting on an implicit vacuum $|\varnothing_{123}\rangle$. Momentum conservation is therefore also encoded. For practical calculation we can think of the annihilators $\alpha$ and $\bar{\alpha}$ as derivatives with respect to the variables $\bar{\alpha}^\dagger$ and $\alpha^\dagger$
\begin{equation}\label{eq:DerivativeRepresentation}
\alpha=\frac{\partial}{\partial \bar{\alpha}^\dagger}\quad\text{ and }\quad\bar{\alpha}=\frac{\partial}{\partial \alpha^\dagger}.
\end{equation}
The equation can be seen as a multidimensional Riccati-type differential equation (see section \ref{subsec:Riccati}). The parameters $\rho^{\,2}$ and $\eta$ will be explained below.

\subsection{Rationale}\label{subsec:Rationale}

The rationale for the equation was explained in \cite{AKHB2013a} based on observations made in \cite{AKHB1988}. The argument can be briefly stated as follows. Consider a two-particle mechanical system with coordinates $t^\mu$ and $b^\mu$ and corresponding canonical momenta $u_\mu$, $d_\mu$. Call these variables ''end-point'' variables.  The center of motion $x^\mu$ and relative coordinates $\xi^\mu$ are defined by
\begin{equation}\label{eq:TopBottomTranscription}
x^\mu={1\over 2}(t^\mu+b^\mu),\quad\xi^\mu={1\over 2}(t^\mu-b^\mu).
\end{equation}
The corresponding canonical conjugate momenta $p_\mu$, $\pi_\mu$ are
\begin{equation}\label{eq:UpDownTranscription}
p^\mu=u^\mu+d^\mu,\quad\pi^\mu=u^\mu-d^\mu.
\end{equation}
In terms of these variables, the mechanical first class constraints of higher spin gauge field theory can be expressed as $p^2\approx0$, $\xi\cdot p\approx0$ and $\pi\cdot p\approx0$. Using equations \eqref{eq:TopBottomTranscription} and \eqref{eq:UpDownTranscription} the constraints become
\begin{equation}\label{eq:Constraints1}
(t-b)\cdot(u-d)\approx0\,,\quad\quad u^2+u\cdot d\approx0\,,\quad\quad d^{\,2}+u\cdot d\approx0.
\end{equation}
Then requiring the endpoints to move with the velocity of light forces the further constraint $u\cdot d\approx0$. This equation is related to one of the Wigner equations defining the continuous spin representations of the Poincar{\'e} group \cite{BargmannWigner1947}. It can also be related to the tracelessness constraints of higher spin theory. The reader is referred to \cite{AKHB2013a} for fuller a discussion of these questions. Using equations \eqref{eq:UpDownTranscription} we have $u\cdot d=(p^2-\pi^2)/4$ so that the constraint can also be written $p^2-\pi^2\approx0$.

The constraint is then applied to the cubic vertex 
\begin{equation}\label{eq:UDequation1}
\sum_{r=1}^3 u_r\cdot d_r\vert V_{123}\rangle=0.
\end{equation}
In order to arrive at the PDE of equation \eqref{eq:UDequation0} we must fix the light-front gauge. We will first do it for free fields and then discuss modifications in the interacting theory. 

\subsection{Light-front gauge fixing}\label{subsec:LFGauge}

The simplest way is to start with the form $p^2-\pi^2\approx0$. Since in the light-front gauge we are on the free field mass shell we have $p^2\approx0$. For $\pi^2$ we have $\pi^2=2(\pi\bar{\pi}-\pi^+\pi^-)$. The light-front gauge for is reached by putting $\pi^+=0$ so what remains is simply $\pi\bar{\pi}\approx0$.

The result can be checked by working from the $u\cdot d\approx0$ form of the constraint. In light-front coordinates we have $u\cdot d=\bar{u}d+u\bar{d}-u^-d^+-u^+d^-$. The gauge is reached by putting $\alpha^+=\alpha^{\dagger+}=0$. This gives for the $+$ components of $u$ and $d$
\begin{align}\label{eq:Uplus}
u^+&=(p^++\pi^+)/2=p^+/2-i(\alpha^+-\alpha^{\dagger+})/2\sqrt2=p^+/2,\\\label{eq:Dplus}
d^+&=(p^++\pi^+)/2=p^+/2-i(\alpha^+-\alpha^{\dagger+})/2\sqrt2=p^+/2.
\end{align}
For the $-$ components we have
\begin{align}\label{eq:Uminus}
u^-&=(p^--\pi^-)/2=p^-/2-i(\alpha^--\alpha^{\dagger-})/2\sqrt2,\\\label{eq:Uminus}
d^-&=(p^--\pi^-)/2=p^-/2-i(\alpha^--\alpha^{\dagger-})/2\sqrt2.\\
\end{align}
The $-$ components of $\alpha$, $\alpha^\dagger$ and $p$ are solved for from the higher spin constraints $\alpha\cdot p\approx0$, $\alpha^\dagger\cdot p\approx0$ and $p^2\approx0$ with the result
\begin{align}
\alpha^-&=\frac{\bar{\alpha}p+\alpha\bar{p}}{p^+},\\
\alpha^{\dagger-}&=\frac{\bar{\alpha}^\dagger p+\alpha^\dagger\bar{p}}{p^+}.\\
p^-&=\frac{p\bar{p}}{p^+}.
\end{align}
We can now calculate $u^-d^+-u^+d^-=p\bar{p}/2$. On the other hand a short calculation shows that $\bar{u}d+u\bar{d}=(p\bar{p}-\pi\bar{\pi})/2$. In this way we reproduce the result $u\cdot d=-\pi\bar{\pi}/2$.

In any way, when the light-front gauge fixed operator $-\pi^2=-2\pi\bar{\pi}$ is expressed in terms of transverse oscillators it becomes
\begin{align}\nonumber
-\pi^2=-2\pi\bar{\pi}&=-2(i/\sqrt2)^2(\alpha^\dagger-\alpha)(\bar{\alpha}^\dagger-\bar\alpha)\\
&=\alpha\bar\alpha+\alpha^\dagger\bar{\alpha}^\dagger-\alpha^\dagger\bar{\alpha}-\alpha\bar{\alpha}^\dagger.
\end{align}
There is a normal ordering issue that we have to resolve. We can choose to normal order either before or after fixing the light-front gauge. Suppose we want to normal order $\alpha\cdot\alpha^\dagger$ in $D$ space-time dimensions. Normal ordering (N.O) first and then light-front gauge fixing (L.F) would yield $\alpha_i^\dagger\alpha_i+D$. On the other hand, first gauge fixing and then normal ordering would yield $\alpha_i^\dagger\alpha_i+D-2$. In four dimensions we parametrize this choice by writing
\begin{equation}\label{eq:PiSquared}
-\pi^2=\alpha\bar\alpha+\alpha^\dagger\bar{\alpha}^\dagger-\alpha^\dagger\bar{\alpha}-\bar{\alpha}^\dagger\alpha-\eta,
\end{equation}
with $\eta=D/2$ in the first case (first N.O then L.F) and $\eta=(D-2)/2$ in the second (first L.F then N.O). The choice will have consequences.

\subsection{Why Riccati?}\label{subsec:Riccati}
We can now see why the equation can be designated as being of Riccati type. A one-dimensional analogue to the $\pi^2$ operator would be
\begin{equation*}
\left(\frac{d}{dx}-x\right)\left(\frac{d}{dx}-x\right).
\end{equation*}
If we let this operator act on a function $e^f$ and equate to zero, we get
\begin{equation*}
f^{\prime\prime}-2xf^\prime+(f^\prime)^2=1-x^2.
\end{equation*}
Then substituting $y=f^\prime$ we get a Riccati-type equation
\begin{equation*}
y^{\prime}-2xy+y^2=1-x^2.
\end{equation*}
Equation \eqref{eq:UDequation0} can be seen as a multidimensional generalization of this simple differential equation.

\subsection{Interacting theory}\label{subsec:Interaction}

Generically we write the vertex as $e^\Delta|\varnothing\rangle$ as in \eqref{eq:CubicVertex}. Now, in determining the equations that govern the form of the $\Delta$-operators we act with light-front Poincar{\'e} generators $g$ on the vertex. Since they are linear operators we then get sums of terms of the form $(g\Delta)e^\Delta|\varnothing\rangle$ equal to zero. Thus, for the cubic vertex it does not really matter if we work with a vertex of the form $e^\Delta|\varnothing\rangle$ or simply $\Delta|\varnothing\rangle$. The equations for $\Delta$ become exactly the same.

This cubic ambiguity is reflected in an observation made in \cite{AKHB2012a} that the restrictions on $\Delta$ from the Poincar{\'e} algebra are very weak, allowing an countable infinite set of vertex operators as listed in \eqref{eq:VertexFunctionsGeneral}. 

The KLT-relations \cite{KawaiLewellenTye1986}, in the field theory limit, relates gravity amplitudes to Yang-Mills amplitudes. In a certain sense gravity tree amplitudes can be considered as squares of Yang-Mills tree amplitudes. In the light-front approach this is explicit for cubic vertices and it can be generalized to arbitrary integer spin. The momentum structure of a pure helicity $\lambda$ cubic vertex being simply
\begin{equation}\label{eq:CubicMomentumStructure}
\Big(\frac{\gamma_1}{\gamma_2\gamma_3}\mathbb{P}\Big)^\lambda.
\end{equation}
This was observed already in \cite{BBL1987} (see formula (22) of that paper). This structure has also been observed by Ananth in \cite{Ananth2012un} using MHV-notation. Indeed, the $\mathbb{P}$ and $\mathbb{\bar{P}}$ and are essentially the same thing as the spinor products $\langle k\,l\rangle$ and $[k\,l]$ respectively (see also comments in \cite{AKHB2012a}).

Anyway, working with a vertex of the form $e^\Delta|\varnothing\rangle$, all pure higher helicity cubic interactions can be generated from the operators $\Delta_{\mathrm{hs}}$ of formula \eqref{eq:CubicVertexDeltasHS} by expanding out the powers in $e^\Delta$. Working with a vertex of the form $\Delta|\varnothing\rangle$ the higher helicity interactions come instead from the operators
\begin{equation}\label{eq:VertexFunctionsHigherHelicity}
\Delta=\kappa^{(k-l)}Y^{r_1\ldots r_ks_1\ldots s_l}\big(\boldsymbol \alpha_{r_k}^\dagger\bar{\boldsymbol \alpha}_{s_l}^\dagger\mathbb{\bar{P}}^{(k-l)}+\bar{\boldsymbol \alpha}_{r_k}^\dagger\boldsymbol \alpha_{s_l}^\dagger\mathbb{P}^{(k-l)}\big),
\end{equation}
with
\begin{equation}\label{eq:YFunctionsHigherHelicity}
Y^{r_1\ldots r_ks_1\ldots s_l}=\frac{\gamma_{s_1}\ldots\gamma_{s_l}}{\gamma_{r_1}\ldots\gamma_{r_k}}.
\end{equation}
Here, the helicity is $\lambda=k-l$. Since for a cubic pure helicity $\lambda$ vertex we must also have the total number of oscillators $k+l=3\lambda$ we get $k=2\lambda$ and $l=\lambda$. For instance for spin $2$ we get the operator
\begin{equation}\label{eq:YFunctionsHigherHelicity}
\frac{\gamma_{s_1}\gamma_{s_2}}{\gamma_{r_1}\gamma_{r_2}\gamma_{r_3}\gamma_{r_4}}\left(\alpha_{r_1}^\dagger\alpha_{r_2}^\dagger\alpha_{r_3}^\dagger\alpha_{r_4}^\dagger\bar\alpha_{s_1}^\dagger\bar\alpha_{s_2}^\dagger\mathbb{\bar{P}}^2+\bar\alpha_{r_1}^\dagger\bar\alpha_{r_2}^\dagger\bar\alpha_{r_3}^\dagger\bar\alpha_{r_4}^\dagger\alpha_{s_1}^\dagger\alpha_{s_2}^\dagger\mathbb{P}^2\right).
\end{equation}
This generalizes to higher helicity in an obvious way. 

An argument for choosing a vertex of the form $e^\Delta|\varnothing\rangle$ rather than $\Delta|\varnothing\rangle$ can therefore hardly be construed from the cubic theory alone. However, once quartic vertices are considered it is probably essential to work with $e^\Delta|\varnothing\rangle$ since the Poincar{\'e} algebra then will contain combinations of the form 
\begin{equation*}
e^{\Delta_{12j}}|\varnothing_{12j}\rangle e^{\Delta_{j34}}|\varnothing_{j34}\rangle,
\end{equation*}
with a contraction over one of the Fock spaces (indexed by $j$ in the qualitative formula above). Then the properties of the exponential function are likely to be crucial. As many authors have commented, the fate of the light-front approach to higher spin is likely to be settled by a calculation of the full quartic vertex -- if it exists.

Now applying the light-front gauge fixed operator $-\pi^2$ from equation \eqref{eq:PiSquared} to the $e^\Delta|\varnothing\rangle$ vertex yields
\begin{equation}\label{eq:UDequation2}
\left(\sum_{r=1}^3(\bar{\alpha}_r\alpha_r-\bar{\alpha}^\dagger_r\alpha_r-\alpha^\dagger_r\bar{\alpha}_r)\Delta+(\bar{\alpha_r}\Delta)(\alpha_r\Delta)+\bar{\alpha}^\dagger_r\alpha^\dagger_r-\eta\right)e^\Delta|\varnothing_{123}\rangle.
\end{equation}

Then there is one more issue to deal with. We started with the operator $p^2-\pi^2$ but concluded that $p^2=0$ on the light-front. But that is for free fields. In the cubic interacting theory we should expect to have instead a term with a factor $\mathbb{P}\mathbb{\bar{P}}$. The simplest such term that is dimensionally correct (same dimension as $p^2$) and symmetric in the field labels is
\begin{equation*}
\sum_{r=1}^3\frac{\mathbb{P}\mathbb{\bar{P}}}{\gamma_r^{\,2}}
\end{equation*}
Adding this term with a parameter $\rho^2$ to \eqref{eq:UDequation2} and equating to zero and dropping $e^\Delta|\varnothing\rangle$ finally yields the PDE recorded above in formula \eqref{eq:UDequation0}.

\section{Consequences of the PDE}\label{sec:Consequences}
Returning now to the vertex operators of \eqref{eq:VertexOperators} everything is fixed by the Poincar{\'e} algebra except the relative numerical coefficients of these operators. We will use the notation $y_{kl}$ (see formula \eqref{eq:VertexOperators}) for the as yet undetermined coefficient of the $Y^{r_1\ldots r_ks_1\ldots s_l}$ term i $\Delta$. For instance the coefficient for the term $Y^{rst}(\alpha_r^\dagger\alpha_s^\dagger\bar{\alpha}_t^\dagger\bar{\mathbb{P}}+\bar{\alpha}_r^\dagger\bar{\alpha}_s^\dagger\alpha_t^\dagger\mathbb{P}$) is $y_{21}$.

\subsection{Fixing the coefficients}\label{subsec:FixingCoefficients}
The procedure is straightforward to check the consequences of the PDE. We will first record the lowest order in oscillators and momenta as parameterized by $k$ and $l$.
\paragraph{$k=0,l=0$:}
Here we get two equations
\begin{align}\label{eq:K0L01}
3y_{11}&=3\eta\\\label{eq:K0L02}
y_{10}^2&=\rho^2
\end{align}
%
\paragraph{$k=1,l=0$:}Here we get the equation
\begin{equation}\label{eq:K1L0}
y_{10}(y_{11}-1)+6y_{21}=0
\end{equation}
We see at once that we can have neither $\eta=1$ nor $\rho=0$ because then $y_{21}=0$. Then we wouldn't have any cubic interactions since this is the coefficient in front of the crucial cubic vertex operator of \eqref{eq:CubicVertexDeltasHS}. As shown above, $\eta=1$ corresponds to normal ordering after light-front gauge fixing. Instead $\eta$ could be treated as a parameter $\not=1$ but I will conveniently choose its value to be $2$ corresponding to normal ordering before light-front gauge fixing.
%
\paragraph{$k=1,l=1$:}Here we get the equation
\begin{equation}\label{eq:K1L1}
y_{11}(y_{11}-2)+12y_{22}=-1
\end{equation}
where the $-1$ comes from the term $-\sum_{r=1}^3\bar{\alpha}^\dagger_r\alpha^\dagger_r$.

The solution to equations \eqref{eq:K0L01}, \eqref{eq:K0L02}, \eqref{eq:K1L0} and \eqref{eq:K1L1} is
\begin{equation}
y_{10}=\rho,\quad y_{11}=2,\quad y_{21}=-\frac{\rho}{6},\quad y_{22}=-\frac{1}{12}.
\end{equation}
On the next $k$ level up we get three equations
\begin{align}
k=2,l=0&: \quad 2y_{20}(y_{11}-1)+9y_{31}+y_{21}y_{10}=0,\\
k=2,l=1&: \quad 3y_{21}(y_{11}-1)+18y_{32}+2y_{22}y_{10}=0,\label{eq:K2L1}\\
k=2,l=2&: \quad 4y_{22}(y_{11}-1)+27y_{33}=0.
\end{align}
Here we have four new coefficients: $y_{20}$ and $y_{31},y_{32},y_{33}$. The equations can be solved in terms of $y_{20}$ with the result
\begin{equation}
y_{31}=\frac{1}{54}\left(\rho^2-12y_{20}\right),\quad y_{32}=\frac{\rho}{27},\quad y_{33}=\frac{1}{81}.
\end{equation}
This pattern continues as can be seen on the next $k$ level up, where we get four equations
\begin{align}\label{eq:K3L}
k=3,l=0&: \quad 3y_{30}(y_{11}-1)+12y_{41}+y_{31}y_{10}+2y_{21}y_{20}=0,\\
k=3,l=1&: \quad 4y_{31}(y_{11}-1)+24y_{42}+2y_{32}y_{10}+4y_{22}y_{20}+2y_{21}^2=0,\\
k=3,l=2&: \quad 5y_{32}(y_{11}-1)+36y_{43}+3y_{33}y_{10}+6y_{21}y_{22}=0,\\
k=3,l=3&: \quad 6y_{33}(y_{11}-1)+48y_{44}+4y_{22}y_{22}=0.
\end{align}
There are now five new coefficients: $y_{30}$ and $y_{41},y_{42},y_{43},y_{44}$ that can be solved for in terms of $y_{30}$ and $y_{20}$. This pattern continues. On level $k$ we get $k+1$ equations for the $k+2$ coefficients $y_{k0}$ and $y_{k+1,1},y_{k+1,2},\ldots y_{k+1,k+1}$ where the $y_{k+1,j}$ coefficients can be solved for in terms of $y_{k,0},y_{k-1,0},\ldots,y_{20}$. 

For the record, here is the solution to the level $k=3$ equations.
\begin{align*}
y_{41}&=\frac{1}{648}\left(-\rho^3+30\rho y_{20}-162 y_{30}\right),\quad y_{42}=-\frac{11}{1296}\left(\rho^2-6y_{20}\right),\\
y_{43}&=-\frac{11\rho}{1296},\quad y_{44}=-\frac{11}{5184}.
\end{align*}
%

\subsection{General structure of the equations}\label{subsec:GeneralStructureEquations}
We can outline the general structure of the equations. Consider level $k$. There are $k+1$ equations parameterized by $l=0,1,\ldots,k$. Each equation has two linear terms. One is $3(k+1)(l+1)y_{k+1,l+1}$ coming from the $\sum_r\bar{\alpha}_r\alpha_r\Delta$ part of the PDE. The coefficients in these terms are determined at this level. The other linear terms are $-(k+l)y_{kl}$ coming from the $\sum_r(-\bar{\alpha}^\dagger_r\alpha_r-\alpha^\dagger_r\bar{\alpha}_r)\Delta$ part of the PDE. The coefficients in these latter terms are already known from the level $k-1$ equations. The rest of the terms are bilinear in already determined coefficients, coming from the $\sum_r(\bar{\alpha}_r\Delta)(\alpha_r\Delta)$ part of the PDE. The equations take the general form
\begin{equation}
y_{k+1,l+1}=-\frac{1}{3(k+1)(l+1)}\big((k+l)y_{kl}(y_{11}-1)+\text{bilinears}\big)
\end{equation}
for $k\geq 2$ and $0\leq l\leq k$. 

The first equation of this set, the one for $y_{k+1,1}$, contains the new undetermined parameter $y_{k0}$. The rest of the coefficients $y_{k+1,l+1}$ with $1\leq l\leq k$ are solved for in terms of the $y_{kl}$. 

The final result is a set of recursive equations that can be solved in terms of the parameters $y_{10}, y_{20}, y_{30}, \ldots$. Putting all the $y_{k0}$ with $k>1$ to zero (which could be seen as a choice of boundary-values for the PDE), all coefficients can be expressed in terms of the parameter $\rho=y_{10}$. The equations for the coefficients $y_{k+1,1}$ then simplify to $3(k+1)y_{k+1,1}=\rho y_{k,1}$ (for $k>1$) with solution
\begin{equation*}
y_{k+1,1}=(-\frac{\rho}{3})^k\frac{1}{(k+1)!}.
\end{equation*}
I haven't been able to derive closed formulas for the rest of the coefficients. However, a numerical study shows that they too drop off rapidly with increasing $k$. Furthermore $y_{kl}\sim \rho^{k-l}$.

\subsection{Off-shell counterterms}\label{subsec:OffshellCounterterms}
The theory considered here is a multi-helicity theory. Therefore even with just cubic vertices, loop diagrams will have vertices connecting fields with different helicities and internal lines will propagate various helicities. The vertices are higher-derivative and therefore potential divergencies will occur of arbitrarily high order. Since there are vertices with arbitrarily high powers of momenta, such divergencies can presumably be matched by corresponding counterterms.

Vertex operator terms with coefficients $y_{k0}$ with $k\leq\lambda_1+\lambda_2+\lambda_3$ will produce $\phi_1\phi_2\phi_3$ interactions with the transverse derivative structure $\mathbb{\bar{P}}^{\lambda_1+\lambda_2+\lambda_3}$ which could serve as off-shell counterterms. It is tempting to put all the $y_{k0}$ (with $k>1$) coefficients to zero. Then all the other coefficients in the vertex are fixed in terms of $y_{10}=\rho$. Even if we do this, we will still generate these kinds of interactions through the vertex operators with coefficient $y_{10}$. The coefficients of these terms, then proportional to $(y_{10})^{\lambda_1+\lambda_2+\lambda_3}$, are however fixed. Switching on the $y_{k0}$ terms again give us off-shell counterterms with adjustable coefficients. So it seems that the PDE fixes the cubic vertex completely at the classical level while still allowing for quantum corrections to be absorbed. However, the quantum properties of a theory like this need to be further studied.

\subsection{A note on gravity cubic counterterms}\label{subsec:GravityCounterterms}
In \cite{AKHB2013b} we studied cubic counterterms for light-front gravity. Apart from the two-loop off-shell term with transverse derivative structure $\mathbb{\bar{P}}^6$ we found an infinite set of on-shell one-loop terms with transverse derivative structure $\mathbb{\bar{P}}^3\mathbb{P}$ differing in their $p^+$ (that is $\gamma$) structure. This is somewhat strange because in a pure helicity $2$ theory we would expect just one such term corresponding a $\bar{2}-2-2$ vertex (where helicity $-2$ is denoted by $\bar{2}$). We explained this as a result of higher helicity cubic interactions leaking into the helicity $2$ calculation. This is indeed so because in a formulation that from the outset contains fields of all helicities (as the present one does) there will be vertices connecting fields of helicities $\bar{4}-3-3$, $\bar{6}-4-4$, $\bar{8}-5-5$, et cetera, all with transverse structure $\mathbb{\bar{P}}^3\mathbb{P}$ but with differing $p^+$ structure.

\subsection{A note on coupling constants and scales}\label{subsec:NoteScales}
In this connection it is interesting to discuss the question of scales in Minkowski higher spin theory. There are actually several places where it is natural to introduce coupling constants and scales. Let us go back to simplest vertex operator encoded in $\Delta_{\rm{hs}}$ (see formula \eqref{eq:CubicVertexDeltasHS}). As is well known this operator suffices to produce all pure higher spin cubic interactions. It comes with a dimensionful constant $\kappa$ of mass dimension $-1$. The cubic vertex $\vert V_{123}\rangle$ itself (see formula \eqref{eq:CubicVertex}) comes with a factor $g/\kappa$ where $g$ is a coupling constant of mass dimension zero. Consequently  spin $1,2,3,\ldots$ cubic interactions gets coupling factors $g,g\kappa,g\kappa^2,\ldots$. Of course, the constants $g$ and $\kappa$ are introduced in a rather ad hoc way, essentially to get the dimensions correct.

However, the PDE considered here naturally introduces at least one new scale into the theory, namely $\rho$ (see formula \eqref{eq:UDequation0}) of mass dimension $-1$. It can be interpreted as relating the dimension of the internal momentum $\pi$ scale to the center of motion momentum $p$ of the underlying two-particle object (see subsection \ref{subsec:Rationale}). As such it can perhaps be seen as a remnant of a string scale in a rigid string limit. Now, the simplest choice is to equate $\rho$ to the $\kappa$ in the factor $g/\kappa$ multiplying $\vert V_{123}\rangle$, but there is no compelling reason to do so. Let us instead treat $\rho$ as a new scale. It then enters as the dimensionful factor of the basic operator $\Delta_{\rm{hs}}$. This follows from the solution to the PDE at level $k=1,l=1$ (see formula \eqref{eq:K1L1}) where we get $y_{21}=-\rho/6$. The sequence of spin $1,2,3,\ldots$ coupling factors then would come out as $g\rho/\kappa,g\rho^2/\kappa,g\rho^2/\kappa,\ldots$. But as we will see, this is not the full story.

We saw in section \ref{subsec:FixingCoefficients} that at each level $k$, new undetermined constants $y_{k0}$
are introduced. These can be seen as boundary data for the PDE but they also need to carry mass dimension of $-k$. The simplest choice would be to take $y_{k0}\sim\rho^k$. We could also consider introducing a new coupling constant $\rho_k$ of mass dimension $-k$ for each and every high spin field such that $y_{k0}=\rho_k$ for $k>1$. The new coupling constant at spin level $s$ would be $\rho_{s-1}$. Alternatively, these new constants $\rho_k$ can be thought of as free counter-term adjusting parameters in a quantum field theory of higher spin.

Concretely, the pure spin 1, 2 and 3 cubic coupling constants then come out as 
\begin{equation}\label{eq:PureCubicVertexCouplings}
-\frac{2g}{3\kappa}\rho,\quad \frac{11g}{27\kappa}(5\rho^2+6\rho_2),\quad -\frac{4g}{243\kappa}(1795\rho^3+7641\rho\rho_2+3537\rho_3)
\end{equation}
respectively, with progressively more complicated expressions for higher spin vertices. 

We can also note the coupling factors in front of some of the example vertices of section \ref{subsec:Spin2Examples} (see formulas \eqref{eq:Grav2ssCaseiia1}, \eqref{eq:Grav2ssCaseiib1} and \eqref{eq:HSs22Caseiib1})
\begin{align}
&\bar2-3-3:\quad\quad \frac{2g}{81\kappa}(431\rho^4+4430\rho^2\rho_2+3552\rho_2^2+6552\rho\rho_3+3672\rho_4)\frac{\gamma_1^2}{\gamma_2^3\gamma_3^3}\mathbb{\bar{P}}^4,\\
&\bar3-2-3:\quad\quad -\frac{76g}{9\kappa}(\rho^2+\rho_2)\frac{\gamma_1^3}{\gamma_2^2\gamma_3^3}\mathbb{\bar{P}}^2,\\
&\bar2-2-3:\quad\quad \frac{g}{81\kappa}(431\rho^3+1998\rho\rho_2+1134\rho_3)\frac{\gamma_1^2}{\gamma_2^2\gamma_3^3}\mathbb{\bar{P}}^3.
\end{align}

It is interesting to compare the non-minimal $\bar2-3-3$ vertex of spin 3 in a spin 2 background with the $\bar2-2-4$ vertex of spin 2 in a spin 4 background as these have the same power of transverse momenta. It turns out that the coupling factor is the same but the $\gamma$-structure is different. Indeed we get
\begin{equation}
\bar2-2-4:\quad\quad \frac{2g}{81\kappa}(431\rho^4+4430\rho^2\rho_2+3552\rho_2^2+6552\rho\rho_3+3672\rho_4)\frac{\gamma_1^2}{\gamma_2^2\gamma_3^4}\mathbb{\bar{P}}^4.
\end{equation}

The form of the coupling factor is a consequence of the combinatorics of picking out the relevant terms corresponding to the helicities of the external states, as is the particular $\gamma$-structure. Both the vertex $\bar2-3-3$ and the vertex $\bar2-2-4$ corresponds to picking out two oscillators $\alpha^\dagger$ and six oscillators $\bar\alpha^\dagger$ but differently indexed. The coupling factor becomes the same but the $\gamma$-structure differs. This is of course a general phenomena in this model and it reflects the basically egalitarian nature of spin.

\subsection{A note on a $\Delta|0\rangle$ vertex as opposed to a $\exp\Delta|\varnothing\rangle$}\label{subsec:NoteDeltaVertex}
As argued above in section \ref{subsec:Interaction} a cubic vertex of the type $\exp(\Delta)|\varnothing\rangle$ is probably implied by higher orders in the interaction. Still it is interesting to see what  would be the consequences or requiring the PDE to hold for a vertex of the type $\Delta|\varnothing\rangle$. In that case the equation becomes
\begin{equation}\label{eq:UDequationDeltaVertex}
\sum_{r=1}^3(\bar{\alpha}_r\alpha_r-\bar{\alpha}^\dagger_r\alpha_r-\alpha^\dagger_r\bar{\alpha}_r)\Delta=\big(-\sum_{r=1}^3\bar{\alpha}^\dagger_r\alpha^\dagger_r+\rho^2\sum_{r=1}^3\frac{\bar{\mathbb{P}}\mathbb{P}}{\gamma_r^2}+3\eta\big)\Delta.
\end{equation}
There are no bilinear terms -- this is a simplification -- but on the other hand, the terms on the right hand side now multiply $\Delta$ -- and that is a complication. In particular the occurrence of the term $-\sum_{r=1}^3\bar{\alpha}^\dagger_r\alpha^\dagger_r\Delta$ complicates matters as we now have to treat terms in $\Delta$ that contain traces $\sum_{r=1}^3\bar{\alpha}^\dagger_r\alpha^\dagger_r$ specifically. 

Consider for instance level $k=2,l=1$ where we get an equation for the crucial $y_{21}Y^{r_1r_2s_1}\alpha_{r_1}^\dagger\alpha_{r_2}^\dagger\bar{\alpha}_{s_1}^\dagger\bar{\mathbb{P}}$ operator. The equation also contains a term coming from $\sum_{r=1}^3\bar{\alpha}_r\alpha_r$ acting on $y_{32}Y^{r_1r_2r_3s_1s_2}\alpha_{r_1}^\dagger\alpha_{r_2}^\dagger\alpha_{r_3}^\dagger\bar{\alpha}_{s_1}^\dagger\bar{\alpha}_{s_2}^\dagger\bar{\mathbb{P}}$ producing $18y_{32}Y^{r_1r_2rs_1}_{\quad\quad\;\;\; r}\alpha_{r_1}^\dagger\alpha_{r_2}^\dagger\bar{\alpha}_{s_1}^\dagger\bar{\mathbb{P}}$ (sum over $r$). These terms are mixed up with a term coming from the right hand side, namely $-\bar{\alpha}^\dagger_r\alpha^\dagger_r y_{10}Y^{r_1}\alpha_{r_1}^\dagger\bar{\mathbb{P}}$. 

The $\gamma$-structure of these latter terms are $1/\gamma_{r_1}$, therefore they add to the terms coming from $Y^{r_1r_2r_3s_1s_2}$ and $Y^{r_1r_2rs_1}_{\quad\quad\;\;\; r}$ with one (more) $r_i,s_i$ index pair contracted. This complication propagates up through the equations and has to be treated at every level. 

Rather than dealing directly with this complication we can change variables according to $\Delta\rightarrow\exp(\Delta)$. Then we get back to our original vertex governed by the equation \eqref{eq:UDequation0}. One can actually see how this problem translates into the equations for the coefficients coming from the $\exp(\Delta)$ vertex. In equation \eqref{eq:K1L0} we have bilinear term $y_{10}y_{11}$ and in \eqref{eq:K2L1} the two bilinear terms $3y_{21}y_{11}$ and $2y_{22}y_{10}$ where the factors $y_{11}$ and $y_{22}$ signal the effect of traces $\sum_{r=1}^3\bar{\alpha}^\dagger_r\alpha^\dagger_r$. 

On general grounds we can argue that if we take the PDE considered here seriously, we get the same functional form of the cubic vertex regardless of whether we write it as $\Delta$ or $\exp(\Delta)$. Consider namely the infinite set of vertex operators
\begin{equation}
y_{kl}Y^{r_1\ldots r_as_1\ldots s_l}\big(\boldsymbol \alpha_{r_k}^\dagger\bar{\boldsymbol \alpha}_{s_l}^\dagger\mathbb{\bar{P}}^{k-l}+\bar{\boldsymbol \alpha}_{r_k}^\dagger\boldsymbol \alpha_{s_l}^\dagger\mathbb{P}^{k-l}\big)
\end{equation}
all of which by themselves satisfies the cubic light-front Poincar{\'e} algebra. Summing them over $k\geq1,0\leq l\leq k$ and denoting the sum by $\Delta$ we get a series ansatz for the PDE in the form of \eqref{eq:UDequation0} corresponding to an $\exp(\Delta)$ vertex. On the other hand we can consider the same series $\Delta$ as an ansatz for  PDE in the form of \eqref{eq:UDequationDeltaVertex} corresponding to a $\Delta$ vertex. We see that what we have is actually the same PDE connected by the change of variables $\Delta\rightarrow\exp(\Delta)$. The solution will be expressed with different sets of coefficients $y_{kl}$.

\section{Conclusion and outlook}\label{sec:Conclusion}
The analysis performed here underlines the urgent need to proceed to an attempt to calculate the full quartic higher spin light-front vertex. Most likely the existence of a quartic vertex will put strong restrictions on the cubic vertices. It is interesting to see if such restrictions are consistent with the restrictions on the cubic vertex imposed by the PDE considered in the present paper, and if the PDE holds for the complete quartic vertex as well.

Work on quartic consistency has been reported in \cite{Taronna2011a} and \cite{DempsterTsulaia2012a} in Poincar{\'e} covariant formulations and for particular interactions in \cite{BoulangerLeclerc2006a,BoulangerSundellLeclerc2008}. Without entering into a detailed review, it seems that the results so far are inconclusive, and the problem merits further study.  

In the case of Yang-Mills models it is known since classic work \cite{Utiyama,OgievetskyPolubarinov1963} and more recent work \cite{BarnichHenneauxTatar1993b} that consistency to the second order in the coupling requires that the structure constants satisfy the Jacobi identity. Requiring consistent deformed non-abelian gauge invariance for spin 1 fields fixes the Yang-Mills Lagrangian uniquely up to the choice of gauge group. The same result can be derived in a light-front formulation, and then the form of the quartic vertex is fixed and the cubic structure constants are again required to obey the Jacobi identities.

Similar results hold for interacting spin 2 fields, eventually leading to the full Einstein theory as investigated by many authors  \cite{Kraichnan1955,Wyss1965,Thirring1961,Feynman1962a,Deser1970,Deser1987,Kibble,MansouriChang,Grensing,StelleWest1980} and generalized to supergravity theories \cite{BoulwareDeserKay1979,MacDowellMansouri1977}. The source of this deformation theoretic approach  to interacting gauge field theories can be traced back to Gupta's work on gravity in \cite{Gupta1952,Gupta1954} as pointed out by Fang and Fronsdal in \cite{FangFronsdal1979} -- a paper that can be seen as the genesis of deformation approaches to higher spin gauge field interactions.

If anything can be learned from the low spin cases, it is that one must have complete control of the cubic vertex in order to approach the higher orders and their back-reaction on the cubic level. In higher spin theory one is immediately hit by the complexity of the theory which forces simplifications, assumptions and special cases -- the effect of which are difficult to fathom. It is hard to escape the impression that flat space higher spin theory still lacks a unifying principle strong enough to control the complexity.

\section{Conventions}\label{sec:Conventions}
Coordinates and momenta are given by
\begin{align}\label{eq:LightFrontCoordinatesMomenta}
x^+&=\frac{1}{\sqrt 2}(x^0+x^3), & x^-&=\frac{1}{\sqrt 2}(x^0-x^3),\\
x&=\frac{1}{\sqrt 2}(x^1+ix^2), & \bar{x}&=\frac{1}{\sqrt 2}(x^1-ix^2),\\
p_+&=\frac{1}{\sqrt 2}(p_0+p_3)=-p, & p_-&=\frac{1}{\sqrt 2}(p_0-p_3)=-p^+,\\
p&=\frac{1}{\sqrt 2}(p_1+BP_2), & \bar{p}&=\frac{1}{\sqrt 2}(p_1-BP_2).
\end{align}
With a mostly-plus Minkowski metric $-+++$, the light-front scalar product becomes
\begin{equation}\label{eq:LightFrontScalarProduct}
\begin{split}
A_\mu B^\mu&=A\bar{B}+\bar{A} B+A_-B^-+A_+B^+\\
&=A\bar{B}+\bar{A} B-A^+B^--A^-B^+
\end{split}
\end{equation}
The transverse oscillators are
\begin{equation}\label{eq:LightFrontOscillators}
\begin{split}
\alpha&=\frac{1}{\sqrt 2}(\alpha_1+i\alpha_2),\hspace{28pt} \bar{\alpha}=\frac{1}{\sqrt 2}(\alpha_1-i\alpha_2),\\
\alpha^\dagger&=\frac{1}{\sqrt 2}(\alpha_1^\dagger+i\alpha_2^\dagger),\hspace{28pt} \bar{\alpha}^\dagger=\frac{1}{\sqrt 2}(\alpha_1^\dagger-i\alpha_2^\dagger).
\end{split}
\end{equation}
They obey commutators
\begin{equation}
[\alpha,\bar{\alpha}^\dagger]=1,\quad\quad [\bar\alpha,\alpha^\dagger]=1.
\end{equation}
%

\section*{Acknowledgment}\label{sec:Acknowledgement}
The present work is based on thinking that goes a long way back in time. I actually did some of the calculations in 1987 but did not trust the results at the time. Much of the rethinking that went into it was done in the very pleasant and creative atmosphere of The Galilie Galileo Institute for Theoretical Physics in Florence during the Higher Spins - workshop in the spring of 2013. In particular I would like to thank Bo Sundborg for deep discussions into higher spin field theory. These discussions indeed go back to the late 1980's and was continued during the workshop, some of them over breakfast at the very nice B\&B Dimorra Nonna Grazie. Hopefully there will be more physics to wring out of those conversations. I would also like to thank Glenn Barnich, Dario Francia, Xavier Bekaert, Per Sundell, Simon Lyakhovich and Alexej Sharapov for discussions at the workshop. In particular it was interesting to meet and talk to Christian Fronsdal. 

I would also like to thank Mirian Tsulaia, Massimo Taronna and Ruslan Metsaev for e-mail discussions during the revision of the manuscript. 

\pagebreak

\end{document}